\documentclass[multphys,vecphys]{svmult}

\usepackage{makeidx}         
\usepackage{graphicx}        
\usepackage{multicol}        
\usepackage[bottom]{footmisc}

\makeindex             

\catcode`\@=11
\def\gsim{\ifmmode{\mathrel{\mathpalette\@versim>}}
    \else{$\mathrel{\mathpalette\@versim>}$}\fi}
\def\lsim{\ifmmode{\mathrel{\mathpalette\@versim<}}
    \else{$\mathrel{\mathpalette\@versim<}$}\fi}
\def\@versim#1#2{\lower 2.9truept \vbox{\baselineskip 0pt \lineskip 
    0.5truept \ialign{$\m@th#1\hfil##\hfil$\crcr#2\crcr\sim\crcr}}}
\catcode`\@=12
\begin{document}

\title*{New HST Views at Old Stellar Systems}


\author{Alvio Renzini}


\institute{INAF -- Osservatorio Astronomico di Padova, Italy}

%
%
\maketitle

\begin{abstract}
HST has recently revealed that many among the most massive globular
clusters harbor multiple stellar populations, and --most
surprisingly-- some of them are extremely helium rich. How these
clusters managed to generate such complex stellar populations, and
what processes let to so dramatic helium enrichment, is
today one of the most exciting puzzles in the astrophysics of stellar
systems.  HST has also been instrumental in demonstrating that both the
bulge of our own Galaxy and that of M31 are dominated by old stellar
populations, coeval to galactic globular clusters. Ultradeep HST
imaging has also demonstrated that a major component of the M31 halo
is metal rich and much younger than old globular clusters.

\end{abstract}

\section{Introduction}
\label{sec:1}
HST has enormously contributed to the study of old resolved stellar
populations, in the Milky Way as well as in nearby galaxies. As this
Conference is meant to celebrate the impact of HST on European
astronomy, it is worth saying that European astronomers have often
played a prime role in the achievements with HST in this field. This
brief review will focus on the the most exciting HST results obtained
by them in recent years, either leading the corresponding projects, or
participating as co-investigators.

\section{Helium-Rich populations in Globular Clusters}
\label{sec:2}
Globular clusters (GC) have always been prime targets for HST. Yet, a wealth
of extremely exciting and unexpected results have been recently obtained with 
ACS, which partly contradict the long-standing view of these objects
as prototypical {\it simple stellar populations} (SSP), i.e. assemblies 
of coeval stars all with the same chemical composition. Early results with HST
indeed confirmed such a view, showing exquisitely narrow sequences in the
color-magnitude diagram (CMD) of the best studied clusters. HST superior 
spatial resolution was instrumental in producing such CMDs, as it allowed
superb photometric accuracy {\it and} proper-motion decontamination from
foreground/background stars (e.g., for the cluster M4\cite{bed01}).
These early results reinforced the notion that sees GCs as 
viable SSP templates.

One exception to this paradigm was known since the early 'seventies,
i.e. $\omega$ Centauri, whose broad RGB indicated that stars are
distributed over a wide range of metallicities. This is emphatically
illustrated by a recent CMD of $\omega$ Cen from a $3\times 3$ ACS
mosaic, shown in Fig. 1\cite{vil07}, with its multiple
turnoffs, broad RGB, and complex HB morphology. Being the most massive
GC in the Galaxy, $\omega$ Cen was not felt as a too
embarrassing exception: thanks to its deeper potential well it may
have retained enriched gas out of which successive stellar generation
were formed. Perhaps, it also started much more massive
than at present, possibly a compact dwarf galaxy in itself.

\begin{figure}
\includegraphics[width=7cm]{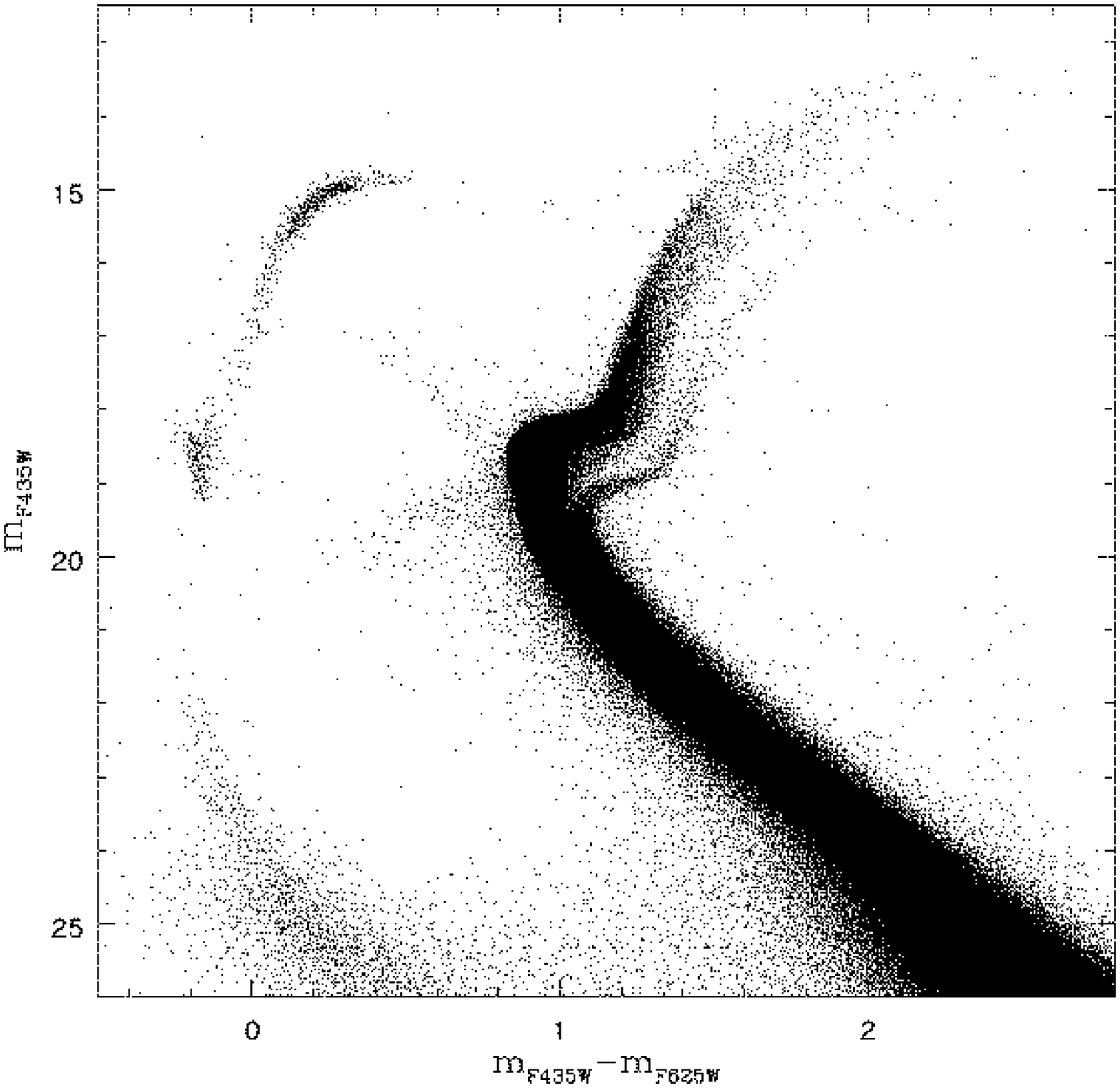}
\hspace{\fill}
\includegraphics[height=6cm, width=4cm]{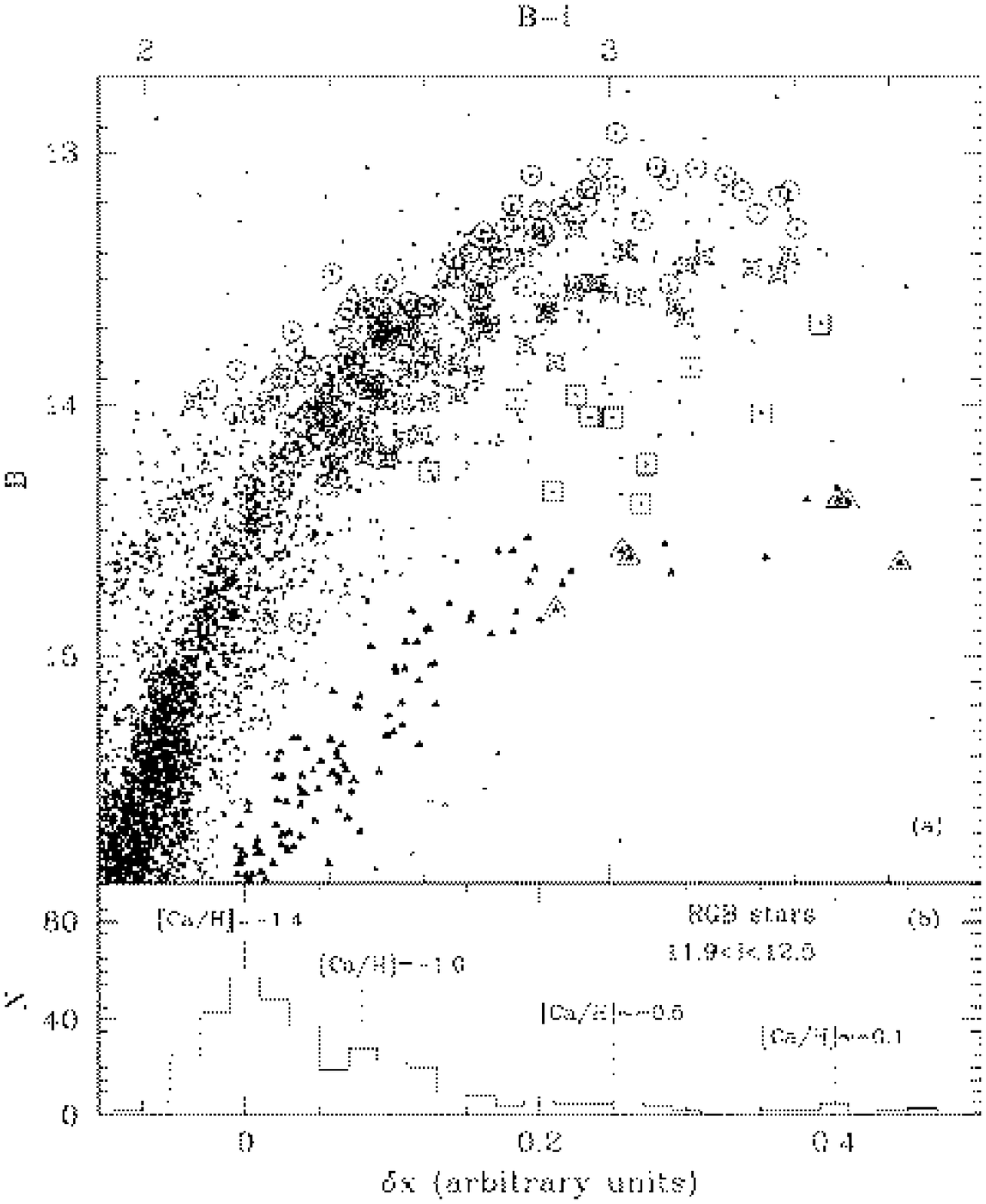} 
\caption{Deep CMD of $\omega$ Cen from a $3\times 3$ ACS
mosaic\cite{vil07} showing all the evolutionary phases from the main
to the white dwarf sequence.  The fine structure of the upper RGB is
shown on the right panel, with the histogram giving the stellar
distribution across a section of the RGB\cite{pan}.}
\label{fig:1}
\end{figure}

\begin{figure}
\centering
\includegraphics[width=6truecm, angle=-90]{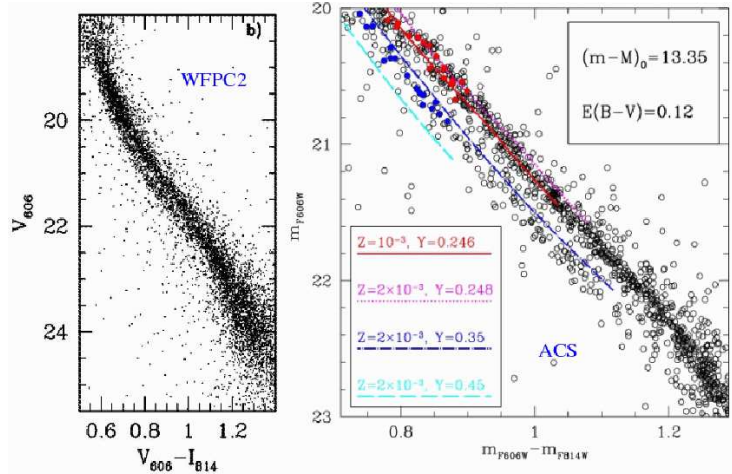}
%
%
\caption{Left panel: the main sequence of $\omega$ Cen splits in two
parallel sequences in this CMD from WFPC2 data\cite{bed04}. Right
panel: the double main sequence of $\omega$ Cen is even better
resolved using ACS data, with overlapped theoretical sequences with
appropriate metallicities and different helium
abundances\cite{pio05}.}
\label{fig:2}       
\end{figure}

A first surprise came from a particularly accurate CMD of $\omega$ Cen
obtained with WFPC2\cite{bed04}, shown here in Fig. 2. The main
sequence (MS) appears in fact split in two parallel sequences,
indicating that at least two distinct star formation episodes had to
take place, rather than a continuous star formation process. A third
distinct population is also evident in Fig. 1, from its faint turnoff
and subgiant branch (SGB) and very red RGB. So, $\omega$ Cen harbors
at least three distinct populations.  Now, the majority of stars in
this cluster are relatively metal poor, as indicated e.g., by the blue
side of the RGB being the most populated (see Fig. 1). One would then
expect that the least populated of the two MSs in Fig. 2 would
correspond to the minority, metal rich population. However, Bedin et
al.\cite{bed04} noted that the metal rich MS should lie to the red of
the metal poor one, instead it lies to the blue!  This is just
contrary to what well understood stellar models predict. Among the
possible solutions of the conundrum Bedin et al.  mention an enhanced
helium abundance up to $Y\sim 0.3$ or more in the metal rich
population, a solution further convincingly explored by
Norris\cite{nor}. That the blue MS is indeed the metal rich one was
then demonstrated spectroscopically using the FLAMES multiobject
spectrograph at the VLT\cite{pio05}, and at this point it became
virtually inescapable to conclude that the cluster contains a minority
population with an helium abundance as high as $Y\simeq 0.35$.

\begin{figure}
\includegraphics[width=6cm]{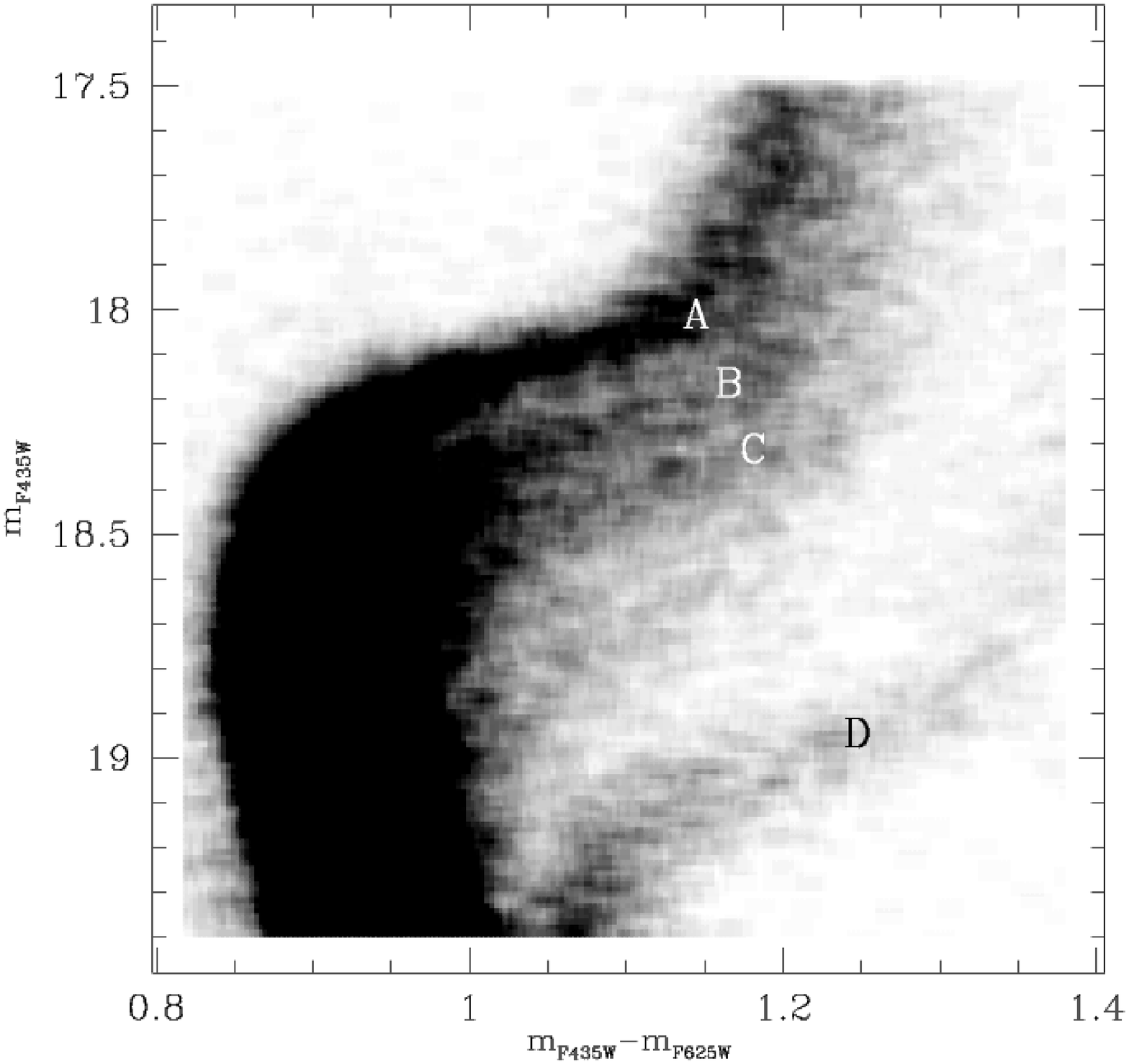}
\hspace{\fill}
\includegraphics[width=6cm]{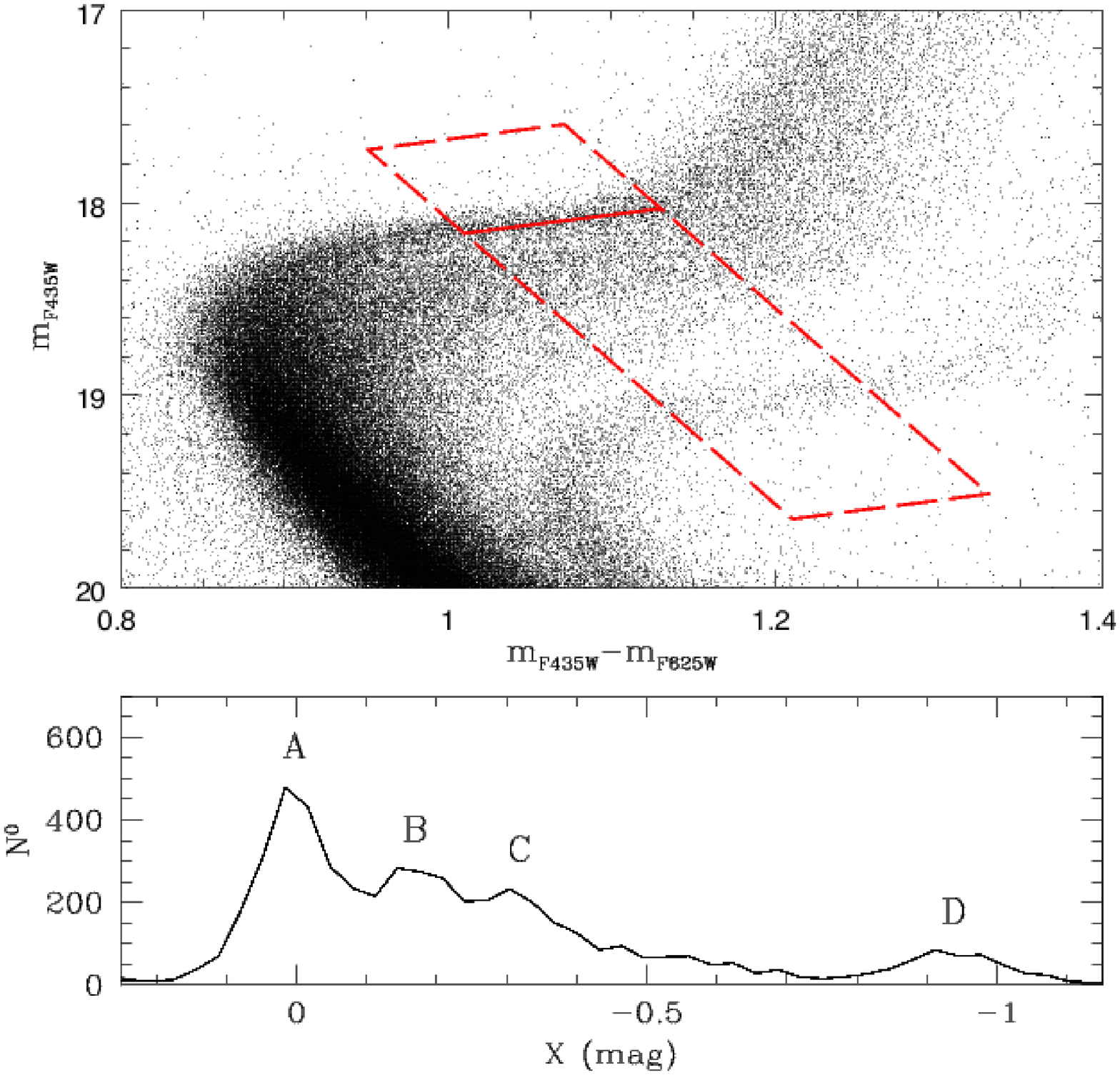}
\caption{Left: the Hess diagram of the turnoff/SGB portion of the
$\omega$ Cen CMD in Fig. 1, showing 4 (perhaps) 5 distinct SGBs.
Right: The same portion of the CMD with indicated the cut used for the 
number counts shown in the lower panel\cite{vil07}.}
\label{fig:3}
\end{figure}

Fig. 3 shows a blow up of the turnoff and SGB region of the CMD in
Fig. 1.  It is clear that at least four distinct populations coexist
(labelled A, B, C, and D), and there is a rather convincing trace of a
fifth one, intermediate between C and D. Indeed, a section of the RGB
shows the presence of five populations, each with a different
metallicity\cite{sol04}. Thus, in $\omega$ Cen one can distinguish at
least 3 MSs (the third being relative to the D population seen in
Fig. 3), 4 or 5 SGBs, 5 RGBs, and a complex, multimodal HB. The real
puzzle is how to connect the various parts of the CMD, recognizing
each of the five MS-SGB-RGB-HB sequences, and estimate age,
metallicity and helium ($t,Z,Y$) for each of the corresponding
populations.

To help composing this puzzle, FLAMES spectroscopy at the VLT was then
undertaken focusing on the SGB
components\cite{sol05}\cite{vil07}.  The conclusions of these two
studies differ somewhat, with one favoring nearly equal ages (within
1-2 Gyr uncertainty) for the five populations, but each with a
distinct helium abundance\cite{sol05}, while the other argues for at
least four age/metallicity groups with a $\sim 30\%$ age range, having
assumed just two helium abundances\cite{vil07}. The $t,Z,Y$
combinations of the five populations remain discrepant from one study
to another (see also\cite{lee}), but only narrowing down these
discrepancies one will understand the formation of
such a complex cluster.

Whereas $\omega$ Cen had for a long time been regarded as a unique
exception, suddenly it started to turn out that it was not  at
all so. Another cluster, NGC 2808 was known for having a multimodal HB,
somehow analog to that of $\omega$ Cen\cite{sos}. Thus, it was
speculated that the multimodal HB of this cluster could also be due to
a multimodal distribution of helium abundances among the cluster
stars\cite{dan}. Rarely speculations receive such a fine observational
confirmation as was the case for this one: Fig. 4 shows the {\it
triple} MS of this cluster, for which no metallicity differences
appear to exist\cite{pio07}. Thus, with 3 distinct MSs, 3 groups of
RGB stars with different [O/Fe] ratios\cite{car}, and an HB made of 4
separated clumps, NGC 2808 hosts at least 3 distinct populations, each with 
different helium, from $Y\simeq 0.24$ up to $Y\simeq 0.37$.

\begin{figure}
\centering
\includegraphics[width=6truecm, angle=-90]{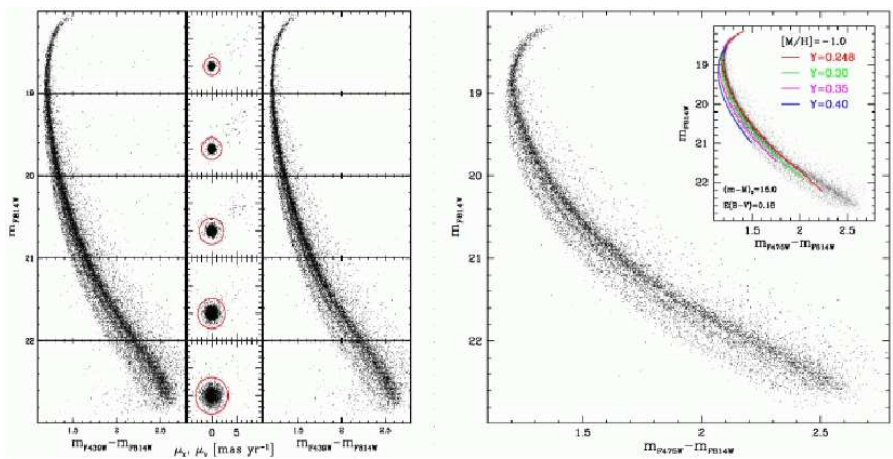}
%
%
\caption{Left panels: the ACS CMD of NGC 2808 before and after proper
motion decontamination, and, in the right panel, the same CMD
corrected for differential reddening, clearly showing its triple main
sequence\cite{pio07}. In the insert theoretical sequences with
different helium abundances are overplotted.}
\label{fig:5}       
\end{figure}

For two other clusters, namely NGC 6388 and NGC 6441, helium-rich
sub-populations have been suggested in order to account for their unique
HB morphology and periods of the RR Lyrae
variables\cite{mol}\cite{cal}.  At this point it soon became apparent
that all 4 GCs with multiple, helium-enhanced populations were
very massive, i.e. among the 11 GCs in the Galaxy which are more
massive than $10^6M_\odot$. Of the remaining most massive clusters, 47
Tuc does not show evidence for multiple populations, but NGC 1851 does
(Piotto et al. in preparation). Among this set of supermassive globulars,
NGC 6715 is being observed in Cycle 15 and NGC 6093, 6388, 7078, and 7089
will be observed in Cycle 16 (PI G. Piotto), in all cases aiming at checking 
whether there is evidence for multiple MSs. So, we shall soon know.

These exciting discoveries still ask more questions than give answers:
\renewcommand{\labelitemi}{$\star$}
\begin{itemize}
\item
How did the most massive GCs manage to form/accrete their multiple stellar
 populations?
\item
Where did the huge amount of helium come from? From 3-8 $M_\odot$  AGB stars? 
Or from where else?
\item
Is $\omega$ Cen the remnant core of a tidally disrupted galaxy?
And if so, what about the other heavyweight GCs?
\item
Are super-helium populations confined to massive GCs? i.e., can we
exclude their presence in massive elliptical galaxies?
\end{itemize}

\section{The Milky Way Bulge and its Globulars}
\label{sec:3}
The Galactic bulge harbors a fair fraction of the total populations of
GCs, and unlike the metal poor GCs in the halo many of them approach
solar metallicity. Observations of two such metal rich GCs taken in
the very first Cycle with WFC2 showed that they are as old as Halo
GCs, demonstrating that the bulge underwent rapid chemical enrichment
and is virtually coeval to the Halo\cite{ort}. Fig. 5 shows a
comparison between the near-IR CMD of a bulge field obtained with
NTT/SOFI, with the CMD of the bulge GC NGC 6528 obtained with
HST/NICMOS\cite{zoc}. The near identity of the two CMDs, and in
particular of the luminosity difference between the HB and the MS
turnoff, ensure that the bulge as a whole is as old as the
cluster, with no trace of an intermediate age population. However,
the bulge CMD was cleaned by the disk contamination in a statistical
fashion, a procedure that might have also removed a trace young
population. A better way of removing the disk contamination was by
picking only HST proper motion bulge members\cite{kon}, which indeed
conclusively demonstrated the virtual absence of an intermediate age
population in the bulge. The question then arose as to whether among bulges
our own is typical or atypical in this respect.

\begin{figure}
\centering
\includegraphics[height=7truecm, angle=-90]{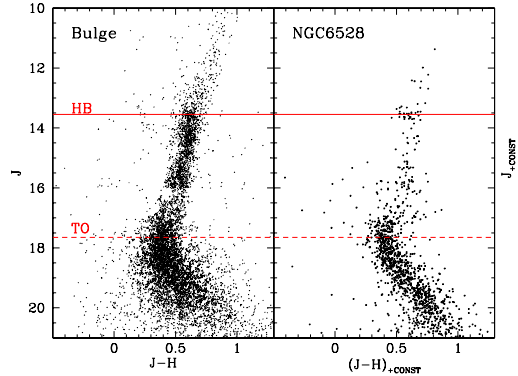}
%
%
\caption{The near-IR CMD of a bulge field from NTT/SOFI data (left) on
the same scale of the HST/NICMOS CMD of the old bulge
globular NGC 6528 (right) \cite{zoc}.}
\label{fig:6}       
\end{figure}

\section{HST Visits to Andromeda}
\label{sec:4} 
Next bulge worth checking is that of M31. There had been early
claims for the M31 bulge being dominated by an intermediate age
population, based on ground based near-IR photometry. This showed an
ubiquitous population of very bright red {\it giants}, then
interpreted as intermediate age AGB stars, but there were good reasons to
doubt such interpretation, given the extreme crowding of the observed
fields. That indeed the apparent bright AGB stars were clumps of
fainter RGB stars was beautifully demonstrated by HST/NICMOS
imaging\cite{m31b}, as illustrated here in Fig. 6. This study also
showed that the near-IR luminosity function of M31 bulge is
indistinguishable from that of the Galactic bulge, hence both bulges
ought to be equally old.

\begin{figure}
\centering
\includegraphics[width=5truecm, angle=-90]{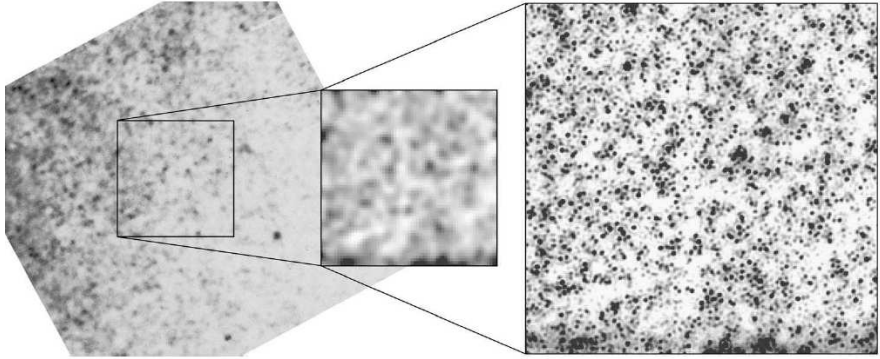}
%
%
\caption{A ground based $K$-band image of a field in the M31 bulge,
and on the right the same field as seen by HST/NICMOS\cite{m31b}. What
appear as individual bright ``stars'' on the ground image, are resolved into
stochastic clumps of many fainter stars on the HST image.}
\label{fig:7}       
\end{figure}

\begin{figure}
\centering
\includegraphics[width=5.7truecm, angle=-90]{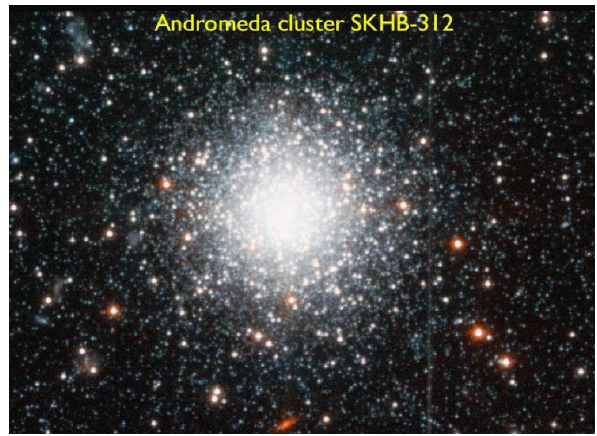}
%
%
\caption{A globular cluster in M31 looks so close in this HST/ACS
image\cite{clu}.}
\label{fig:8}       
\end{figure}

With the advent of ACS it became affordable trying to reach the old MS
turnoff in M31. This was first accomplished on an inner halo field,
$\sim 11$ kpc from the nucleus on the minor axis, investing some 120
HST orbits\cite{bro}. Surprisingly, the result was that, besides the
old/metal poor population, the field includes also an intermediate age
($\sim 7$ Gyr old), metal rich population making up to 40\% of the
total.  Subsequent HST/ACS projects by the same team probed a disk, a
stream, and more outer halo fields in the attempt of mapping the star
formation/assembly history of our companion spiral that looks so
similar, yet turns out to be so different from ours. There is no room
left to make justice of these HST results, and to
illustrate the power of HST/ACS should suffice to show in Fig. 7 the image of
a GC in M31\cite{clu}.

With the Milky Way and the M31 bulges being dominated by stars at
least $\sim 10$ Gyr old, this is like saying that these
bulges had to form the bulk of their stars at $z\gsim 1.5$, and evolve
passively since then. If so, we should see passive galaxies at such
high redshifts in deep spectroscopic surveys, and indeed we do, and
once more HST has been instrumental in reveling their {\it
spheroidal} morphology\cite{cim}\cite{dad}.

In summary, there has been a very strong and successful use of HST by
European astronomers in the field of resolved stellar populations
(globular clusters, MW \& M31 bulges and more) as well as on high
redshift galaxies. It is also important to stress that the HST+VLT
synergy has been very effectively exploited for many programs.
Much more is now expected to come in the final years of HST 
(2008-2014), hopefully  with a telescope more  powerful than ever
(thanks to WFC3/COS/ACS/STIS/NICMOS/FGS).

I would like to thank again Duccio Macchetto, for having invited me to
this exciting conference, and for having done so much for HST and for
attracting the European astronomers to its scientific use.

%

\begin{thebibliography}{99.}
%
%
%

\bibitem{bed01} Bedin, L.R., Anderson, J., King, I.R., Piotto, G. 2001, ApJ, 
                560, L75
\bibitem{vil07} Villanova, S., Piotto, G., King, I.R., Anderson, J., Bedin, 
                L.R., Gratton, R.G., Cassisi, S., Momany, Y., Bellini, A.,
                Cool, A.M., et al. 2007, ApJ, 663, 296
\bibitem{pan}   Pancino, E., Ferraro, F.R., Bellazzini, M., Piotto, G.,
                Zoccali, M. 2000, ApJ, 534, L83 
\bibitem{bed04} Bedin, L.R.,  Piotto, G., Anderson, J., Cassisi, S., King, 
                I.R., Momany, Y., Carraro, G. 2004, ApJ 605, L125
\bibitem{nor}   Norris, J.E. 2004, ApJ, 612, L25
\bibitem{pio05} Piotto, G., Villanova, S., Bedin, L.R., Gratton, R. Cassisi, 
                S., Momany, Y., Recio-Blanco, A., Lucatello, S., Anderson, J.
                et al. 2005, ApJ, 621, 777
\bibitem{sol04} Sollima, A., Ferraro, F.R., Pancino, E., Bellazzini, M. 2005, 
                MNRAS, 357, 265
\bibitem{sol05}	Sollima, A., Pancino, E., Ferraro, F.R., Bellazzini, M., 
                Straniero, O., Pasquini, L. 2005, ApJ, 634, 332
\bibitem{lee}   Lee, Y.-W., Joo, S.-J., Han, S.-I., Chung, C., et 
                al. 2005, ApJ, 621, L57
\bibitem{sos}   Sosin, C., Dorman, B., Djorgovski, S.G., Piotto, G., et al. 
                1997, ApJ, 480, L35
\bibitem{dan}   D'Antona, F., \& Caloi, V. 2004, ApJ, 611, 871
\bibitem{pio07} Piotto, G., Bedin, L.R., Anderson, J., King, I.R., Cassisi, S.,
                Milone, A. P., Villanova, S., Pietrinferni, A., Renzini, A. 
                2007, ApJ, 661, L53
\bibitem{car}   Carretta, E., Bragaglia, A., Gratton, R.G., Leone, F., 
                Recio-Blanco, A., Lucatello, S. 2006, A\&A, 450, 523
\bibitem{mol}   M\"ohler, S., \& Sweigart, A.V. 2006, A\&A, 455, 943
\bibitem{cal}   Caloi, V., \& D'Antona, F. 2007, A\&A, 463, 949
\bibitem{ort}   Ortolani, S., Renzini, A., Gilmozzi, R., Marconi, G., Barbuy, 
                B., Bica, E., Rich, R.M.1995, Nature, 377, 701
\bibitem{zoc}   Zoccali, M., Renzini, A., Ortolani, S., Greggio, L., Saviane, 
                I., Cassisi, S., Rejkuba, M., Barbuy, B., Rich, R.M., Bica, E.
                2003, A\&A, 399, 931
\bibitem{kon}   Kuijken, K., \& Rich, R.M. 2002, AJ, 123, 2054
\bibitem{m31b}  Stephens, A.W., Frogel, J.A., DePoy, D.L., Freedman, W., 
                Gallart, C., Jablonka, P., Renzini, A., Rich, R.M., Davies, R.
                2003, AJ, 125, 2473
\bibitem{bro}   Brown, T.M., Ferguson, H.C., Smith, E., Kimble, R.A.,
                Sweigart, A.V., Renzini, A., Rich, R.M., VandenBerg, D.A. 
                2003, ApJ, 592, L17
\bibitem{clu}   Brown, T.M., et al. 2004, ApJ 613, L125
\bibitem{cim}   Cimatti, A., Daddi, E., Renzini, A., Cassata, P., Vanzella, E.,
                Pozzetti, L., Cristiani, S., Fontana, A., Rodighiero, G.,
                et al. 2004, Nature, 430, 184
\bibitem{dad}   Daddi, E., Renzini, A., Pirzkal, N., Cimatti, A., Malhotra, S.,
                Stiavelli, M., Xu, C., Pasquali, A., Rhoads, J.E., Brusa, M. et
                al. 2005, ApJ, 626, 680  

\end{thebibliography}
%



\printindex
\end{document}